# MEMORY ASSESSMENT OF VERSATILE VIDEO CODING

*Arthur Cerveira, Luciano Agostini, Bruno Zatt*
Video Technology Group (ViTech) - PPGC - CDTec
Federal University of Pelotas (UFPel), Pelotas, Brazil
{aacerveira, agostini, bzatt}@inf.ufpel.edu.br

*Felipe Sampaio*
Federal Institute of Rio Grande do Sul (IFRS)
Campus Farroupilha, Farroupilha, Brazil
felipe.sampaio@farroupilha.ifrs.edu.br

**ABSTRACT**

This paper presents a memory assessment of the next-generation Versatile Video Coding (VVC). The memory analyses are performed adopting as a baseline the state-of-the-art High-Efficiency Video Coding (HEVC). The goal is to offer insights and observations of how critical the memory requirements of VVC are aggravated, compared to HEVC. The adopted methodology consists of two sets of experiments: (1) an overall memory profiling and (2) an inter-prediction specific memory analysis. The results obtained in the memory profiling show that VVC access up to 13.4x more memory than HEVC. Moreover, the inter-prediction module remains (as in HEVC) the most resource-intensive operation in the encoder: 60%-90% of the memory requirements. The inter-prediction specific analysis demonstrates that VVC requires up to 5.3x more memory accesses than HEVC. Furthermore, our analysis indicates that up to 23% of such growth is due to VVC novel-CU sizes (larger than 64x64).

## 1. INTRODUCTION

Recently, there has been an increase in the amount of video data volume, primarily due to the emergence of various streaming services on the internet and devices capable of reproducing such media. Storing and transferring digital video data becomes more challenging as Ultra High-Definition (UHD) and Virtual Reality (VR) get even more popular among the general public. This increase creates a higher demand for improved efficiency in video encoders. Video coding standards employ complex and memory-intensive algorithms [1], which become a problem since the memory represents a large portion of the energy consumption in a computing system.

In 2017, a joint call for proposals (CfP) was issued by the Joint Video Experts Team (JVET)[2]. The goal of this CfP was to gather video coding tools with greater compression efficiency than the High Efficiency Video Coding (HEVC) state-of-the-art standard [2]. With the responses to the CfP, JVET started the standardization project for the Versatile Video Coding (VVC) in 2018 [3]. The novelties proposed by next-generation video encoders tend to aggravate the memory challenges, which rely on more complex coding tools to achieve better coding efficiency. Therefore, *there is a strong need for proper evaluations to measure the impact of the novel tools inserted by the next-generation video encoders concerning memory-related aspects.*

Considering the VVC, there are only a few published studies analyzing its novel coding tools, especially when it comes to memory-related topics. In [4]–[7], the VVC coding efficiency and computational complexity are evaluated and compared to HEVC and AV1 [8] codecs. None of these works exploit memory aspects in their analysis. The high complexity of VVC is used as motivation for fast decision schemes at several levels: coding tree unit partitioning level [9]–[13] and prediction level [13]–[15]. However, their techniques do not consider requirements like memory bandwidth and access patterns.

Therefore, the main goal of this work is to analyze the memory accesses of VVC novel coding tools and compare the results with its predecessor, the HEVC standard. As a result, we aim at tracing promise insights and research perspectives in terms of memory optimization to enable energy-efficient VVC application design.

As the **main contributions** of this work, VVC and HEVC memory analyses were carried out through two different perspectives. HM [16] (HEVC) and VTM [17] (VVC) video coding test models were employed for the experiments.

- *Memory profiling (Section 4)*: which leverages a profiling tool to analyze the overall memory requirements, as well as the memory breakdown considering each video encoding module;
- *Inter-prediction memory analysis (Section 5):* which analyzes in detail the most memory-intensive coding module, the inter-frame prediction. The memory accesses are evaluated from an overall perspective and towards each processed block size by the VVC encoder.

---

[2] ITU-T Video Coding Experts Group (VCEG) and the ISO/IEC Moving Picture Experts Group (MPEG).

## 2. VVC CODING STRUCTURES

The main innovation of VVC exploited in our analysis is the more flexible frame partitioning. As in HEVC, the maximum block size is represented by a coding tree unit (CTU), which can be recursively split into coding units (CUs). The VVC defines the CTU size as 128x128, instead of the default 64x64 CTU at HEVC. Additionally, VVC employs a partitioning scheme based on a quadtree with a nested multi-type tree (QT-MTT) [18], composed of binary and ternary splits, which replaces the concepts of partition types defined at prediction unit level adopted by the HEVC standard [19]. Fig. 1 illustrates the QT-MTT partitioning structure defined by VVC. Starting at the CTU level, the QT-MTT strategy defines that, when a leaf node is achieved in the initial quadtree partitioning, the CU can be further divided by adopting combined binary/ternary tree splitting. In the example of Fig. 1, the a, b, c, and d (top-left CUs at CTU - Fig. 1a; and leftmost nodes at decision tree - Fig. 1b) are achieved after two splits in the initial quadtree, followed by the combination of a ternary/binary divisions.

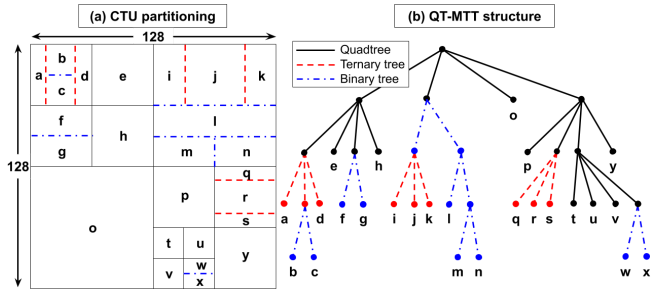

**Fig. 1.** Examples of (a) CTU partitioning and (b) QT-MTT structure defined in VVC.

The higher flexibility of QT-MTT (VVC) in comparison with the adopted quadtree plus PU partitions (HEVC) is illustrated in Fig 2, where the block sizes allowed by each strategy are highlighted in the 2D maps correlating vertical and horizontal dimensions. From an overall perspective, VVC supports 27 different block sizes, whereas HEVC can exploit 20 sizes. Considering the asymmetric partitions, the QT-MTT structure of VVC allows 22 possibilities, which enables a more flexible adaptation of the encoding process in specific frame regions (like frame boundaries) [20]. Furthermore, the 128x128 CTU size supported by VVC enables three novel large size CUs: 128x128 (symmetric), 128x64, and 64x128. Although the use of these large CU sizes enables better compression efficiency (mainly when encoding UHD videos), it increases the computational complexity, since the VVC encoder should analyze more possibilities. The impact of these three novel CU sizes of VVC in terms of memory access is analyzed in Section 5.

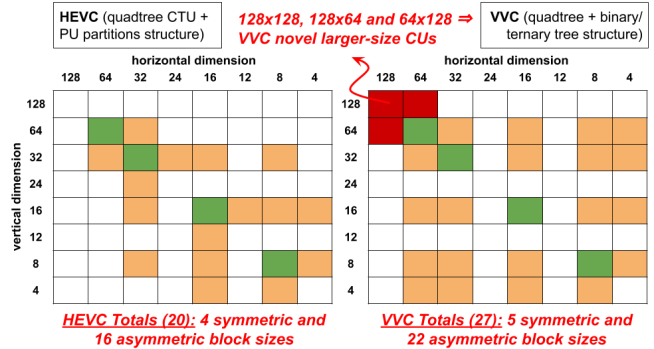

**Fig. 2.** Allowed block sizes of CTU partitioning schemes.

## 3. EXPERIMENTAL SETUP

The VVC memory was analyzed and compared to its predecessor HEVC. Two sets of experiments were performed as a way to measure the memory accesses: (1) an overall memory profiling (Section 4) and (2) a specific inter-frame prediction memory analysis (Section 5). The adopted methodologies for both sets of experiments are depicted in the flowchart of Fig. 3 and described as follows.

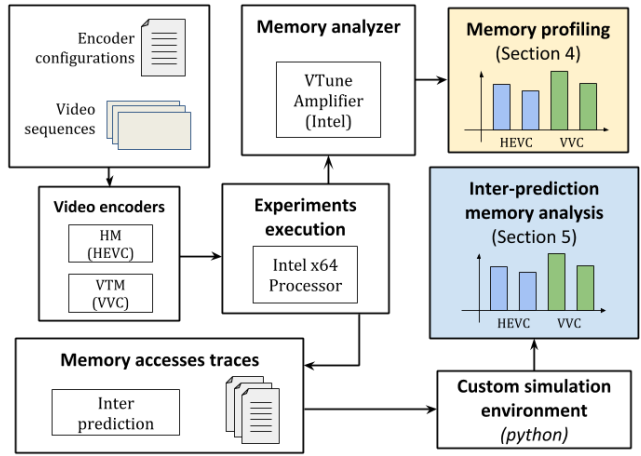

**Fig. 3.** Adopted experimental setup for VVC memory assessment.

The VTM 8.0 was used as a reference test model for VVC evaluation, whereas HM 16.18 was used for HEVC analysis. All analyses were performed according to the common test conditions adopted by the video coding community [21]. Two encoder configurations were selected: Low Delay and Random Access. For the inter-frame prediction analysis, eight video sequences were chosen: four HD1080p (1920x1080 - *BQTerrace*, *BasketballDrive*, *Cactus,* and *ParkScene*), two 2K (2560x1600 - *PeopleOnStreet* and *Traffic*) and two UHD 4K (3840x2160 - *Campfire* and *TrafficFlow*). For the memory profiling, the four HD1080p videos were chosen. The experiments were performed in the first 17 frames of videos. The search range was set as 96 (VTM default value). Higher SR values were considered as well, but there was no significant disparity in the ratio between the encoders.

The overall memory profiling was performed using the Intel® VTune™ Amplifier profiling tool [22] to monitor the volume of memory loads (read accesses) and stores (write accesses) of each module, not considering the influence of a particular processor cache organization. For the inter-frame prediction memory analysis, the HM and VTM software applications were extended to generate memory trace files detailing the accesses to the candidate blocks and their required volume of fetched data. A custom simulator was developed to read these memory traces, extract the information, and calculate the required memory bandwidth for the inter-prediction for each analyzed case.

## 4. MEMORY PROFILING

This section presents the performed memory profiling, which measured the difference between the total volume of memory accesses by VVC compared to HEVC (*Analysis-1* of Section 4.1), as well as the memory breakdown of the memory access by each encoding module (*Analysis-2* of Section 4.2).

### 4.1. Analysis-1: Overall encoder memory accesses

Fig. 4 shows the memory access profiling of the entire executions of HEVC and VVC encoder applications (HM and VTM). The presented results are the sum of all read and write accesses captured by the profiling tool. Moreover, the memory accesses of HM are normalized to the VVC executions, since we are interested in the ratio between them. In all analyzed cases, the VTM executions required significant higher memory accesses indexes: 7.4x-9.1x higher for Low Delay and 12.1x-13.4x higher for Random Access. Although the memory requirements vary according to video content (like texture and motion properties), the ratio between VVC and HEVC results remains similar.

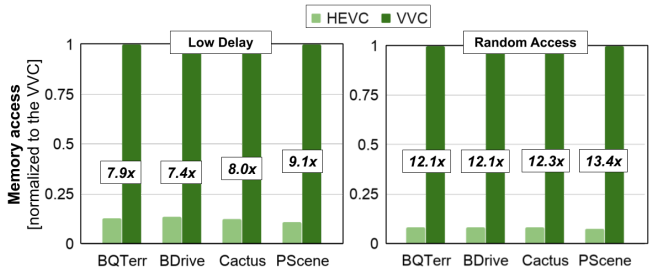

**Fig. 4.** Overall encoder memory access analysis using Intel VTune Amplifier profiling tool.

*Insights from Analysis-1:* Even though this comparison may be affected due to different implementation issues of VTM and HM, which may impact the results reliability, *the significant memory access increasing of VTM (up to 13x more than HM) is an important hint for researchers to go through detailed evaluations in memory-related topics of VVC.*

### 4.2. Analysis-2: Memory accesses profiling per encoding module

Fig. 5 depicts the memory accesses distribution among the different encoding modules of VVC, in comparison with HEVC. The considered encoding modules were: inter-prediction, divided into integer prediction (INTER-I) and fractional prediction (INTER-F), intra-prediction (INTRA), transforms and quantization (T/Q), entropy coding (ENTROP) and filters (FILTERS). The remaining parts were grouped into the "OTHERS", which includes tasks like encoder control and current frames management.

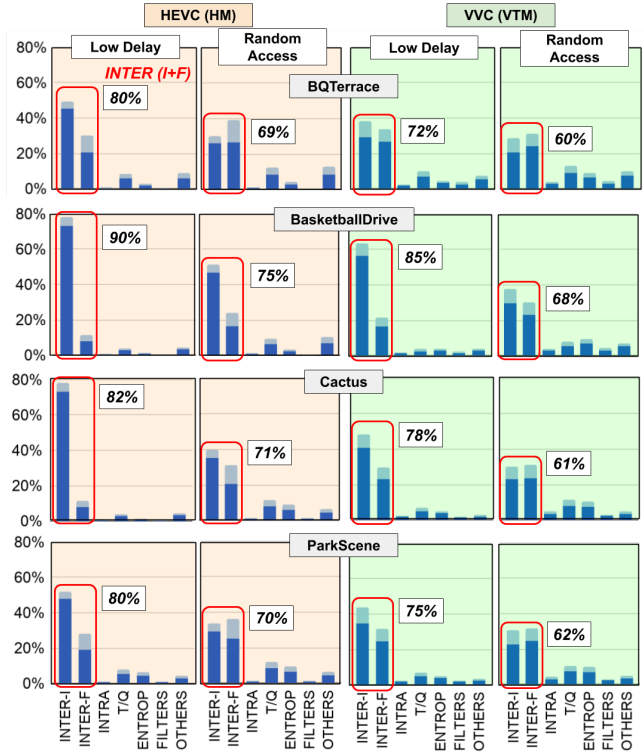

**Fig. 5.** Memory access breakdown of HEVC and VVC encoding modules.

In all analyzed cases, the inter-prediction was responsible for the most memory accesses: from 60%-69% (*BQTerrace*) to 85%-90% (*BasketballDrive*) in VTM and HM, respectively. This variation is due to the video content properties since both videos have the same resolution: *BasketballDrive* presents, in the encoded frames, a high temporal complexity (motion properties). Another important aspect is the representativity of the fractional inter-prediction, which is responsible for 12%-39% of all memory accesses. When analyzing the VTM results, the inter-prediction is still the most memory consuming module, presenting a slightly lower representativity when comparing with HM. The VVC novelties include coding tools for all encoding modules, like intra-prediction, transforms and quantization, and filtering operations, which increases their representativity concerning the memory requirements.

***Insights from Analysis-2:*** *Inter-frame prediction remains the most critical bottleneck at VVC (representing up to 85% of the total encoding memory accesses). Since the overall memory requirements are significantly higher for VVC (up to 13x, as presented in Analysis-1), specific memory evaluations and optimizations in the inter-prediction step are strongly needed to enable memory-efficient VVC encoding.*

## 5. INTER-PREDICTION SPECIFIC MEMORY ANALYSIS

The inter-frame prediction is the most resource-intensive operation in both HEVC and VVC (as discussed in the analyses of the previous section). Considering this, a specific analysis was performed as a way to measure the increased VVC memory requirements in comparison with HEVC. We organize the discussion in two parts: overall perspective (*Analysis-3* of Section 5.1) and CU-size based perspective (*Analysis-4* of Section 5.2).

### 5.1. Analysis-3: Inter-prediction overall memory accesses

Fig. 6 presents the evaluation of the memory accesses specifically for the inter-prediction step. These analyses show that the VVC memory accesses, on average, are 3.5 times higher than the HEVC. The highest difference identified between the encoders was 4.7x (Low Delay) and 5.3x (Random Access), found in the *Campfire* video sequence. The lowest difference was 2.6x in both configurations, found in the *BQTerrace* video sequence.

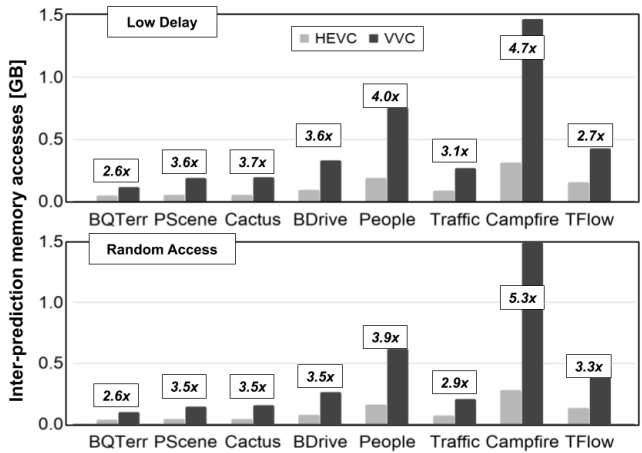

**Fig. 6.** Memory accesses in inter-frame prediction for Low Delay and Random Access encoder configuration.

The discrepancy between the encoders does not ascribe to the total amount of memory accessed during the encoding of each video sequence. For example, even though *PeopleOnStreet* accesses an absolute memory volume ~2x higher than *BasketballDrive*, the memory overheads of VVC remain close: 3.6x and 4.0x for Low Delay, and 3.5x and 3.9x for Random Access, respectively. Furthermore, when analyzing the behavior between Low Delay and Random Access configurations, the ratios between the accessed memory data of each encoder are similar.

***Insights from Analysis-3:*** *The novelties of VVC, like the more flexible QT-MTT partitioning scheme (explained in Section 2), lead to increased memory requirements for inter-frame prediction, which can reach 5.8x of access overhead.*

### 5.2. Analysis-4: Inter-prediction memory accesses per CU size

Fig. 7 presents a graph displaying how much memory is accessed by the novel block sizes introduced on the VVC encoder: 128x128, 128x64, 64x128 (as discussed in Section 2). The inter-frame prediction memory accesses lead by the processing of VVC-novel block sizes represent from 4.1% (*PeopleOnStreet*) to 23.3% (*TrafficFlow*), considering the overall accesses during this prediction step. This range of representativity is strongly related to the video characteristics since tested videos with the same resolution have different results in this analysis: for the HD videos, the large-size VVC CUs represent from 9.0% to 20.7%.

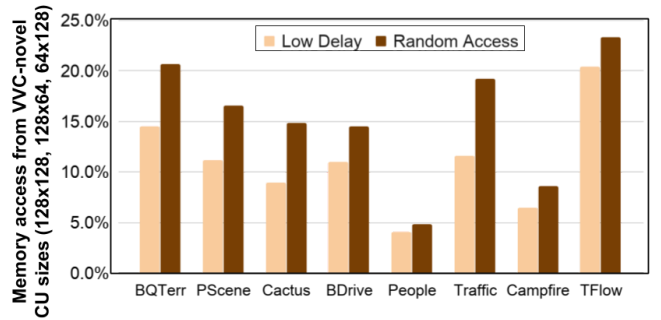

**Fig. 7.** Analysis of memory accesses related to VVC-novel block sizes (128x128, 128x64 and 64x128).

***Insights from Analysis-4:*** *The insertion of large-size CUs (larger than 64x64 - maximum at HEVC) at VVC may represent a significant memory overhead for inter-prediction step (up to 23%). Furthermore, this overhead behavior varies according to the input sequence, and it is strongly dependent on the video content characteristics.*

## 6. CONCLUSIONS

As the main contributions, this paper presents two sets of experiments concerning VVC memory accesses and HEVC memory accesses. The first experiment consists of VVC-overall memory analysis — the second consists of an inter-frame prediction specific analysis. Some of the insights traced from the overall memory analysis include a significant memory access increase in VVC compared to

HEVC, as well as the inter-frame prediction remaining the most critical bottleneck in the encoding process. In the inter-prediction analysis, we observed that the novelties introduced in VVC, such as the QT-MTT partitioning scheme and insertion of large-size CUs, are the responsible for most of the increased memory requirements for this module during the encoding process. Considering all aforementioned insights from our analysis, we conclude that there is an open research gap on minimizing the aggravated VVC memory bottleneck in order to enable energy-efficient next-generation video encoding applications.